\documentclass[superscriptaddress,showpacs,twocolumn,prl]{revtex4}

\usepackage{amsmath}
\usepackage{amssymb}
\usepackage{dsfont}
\usepackage{color}
\usepackage{hyperref}
\usepackage{graphicx}
\usepackage{epstopdf}
\usepackage{soul}
\usepackage{color}

\begin{document}
\title{Modulation-assisted tunneling in laser-fabricated photonic Wannier-Stark ladders}

\author{Sebabrata Mukherjee}
\email{snm32@hw.ac.uk}
\affiliation{SUPA, Institute of Photonics and Quantum Sciences, Heriot-Watt University, Edinburgh, EH14 4AS, United Kingdom}
\author{Alexander Spracklen}
\affiliation{SUPA, Institute of Photonics and Quantum Sciences, Heriot-Watt University, Edinburgh, EH14 4AS, United Kingdom}
\author{Debaditya Choudhury}
\affiliation{SUPA, Institute of Photonics and Quantum Sciences, Heriot-Watt University, Edinburgh, EH14 4AS, United Kingdom}
\author{Nathan Goldman}
\affiliation{Center for Nonlinear Phenomena and Complex Systems, Universit\'e Libre de Bruxelles, CP 231, Campus Plaine, B-1050 Brussels, Belgium}
\author{Patrik \"Ohberg}
\affiliation{SUPA, Institute of Photonics and Quantum Sciences, Heriot-Watt University, Edinburgh, EH14 4AS, United Kingdom}
\author{Erika Andersson}
\affiliation{SUPA, Institute of Photonics and Quantum Sciences, Heriot-Watt University, Edinburgh, EH14 4AS, United Kingdom}
\author{Robert R. Thomson}
\affiliation{SUPA, Institute of Photonics and Quantum Sciences, Heriot-Watt University, Edinburgh, EH14 4AS, United Kingdom}

\begin{abstract}
We observe Wannier-Stark localization in curved photonic lattices, realized using arrays of evanescently coupled optical waveguides. By correctly tuning the strength of inter-site coupling in the lattice, we observe that Wannier-Stark states become increasingly localized, and eventually fully localized to one site, as the curvature of the lattice is increased. We then demonstrate that tunneling can be successfully restored in the lattice by applying a sinusoidal modulation to the lattice position, an effect that is a direct analogue of photon-assisted tunneling. This precise tuning of the tunneling matrix elements, through laser-fabricated on-site modulations, opens a novel route for the creation of gauge fields  in photonic lattices.
\end{abstract}

\maketitle

{\it Introduction} Quantum matter or light propagating in engineered lattices offer versatile platforms for the quantum simulation of new states of matter, such as topological phases \cite{Carusotto:2013gh,Rechtsman:2013,Hafezi:2011dt,Bloch:2008gl,Goldman:2014bv}. This approach relies on novel technologies allowing for the tuning of microscopic parameters that characterize lattice models of interest, such as on-site interactions and tunneling matrix elements. In fact, generating complex tunneling matrix elements can potentially induce non-trivial gauge structures in the lattice, e.g. artificial magnetic fields and non-Abelian gauge potentials \cite{Carusotto:2013gh,Dalibard2011,Goldman:2014bv}, offering an interesting route for quantum simulation \cite{Bloch:2012jy}. In this context, photon-assisted tunneling, a powerful method by which tunneling can be controlled in lattice systems, has been recently exploited in cold gases \cite{Ruostekoski:2002,Jaksch2003,Eckardt2005,Eckardt:2007epl,Gerbier:2010,Kolovsky:2011,Goldman:2015pra} and ion traps \cite{Bermudez:2011prl,Bermudez:2012njp}; this led to the experimental realization of the Hofstadter model \cite{Aidelsburger:2013,Ketterle:2013} and to the detection of the topological Chern number \cite{Aidelsburger:2014hm} with cold atomic gases. 

The photon-assisted-tunneling method relies on two main ingredients \cite{Eckardt2005,Bermudez:2011prl,Kolovsky:2011,Goldman:2015pra}: (a) an artificial electric field generating a large energy offset $\Delta$ between neighboring sites, hence inhibiting the bare hopping, and (b) a time-modulation of the on-site energy, whose frequency  is resonant with respect to the static offset $\omega = \Delta / \hbar$; this restores the tunneling in an efficient and tunable manner. In this Letter, we experimentally demonstrate the realization of modulation-assisted tunneling, an analogue of photon-assisted-tunneling, in arrays of coupled optical waveguides.

The transport of light in a system of coupled optical waveguides, a photonic lattice, can be described by a Schr\"odinger-like equation. As a result, photonic lattices can be used to observe phenomena known from solid state physics. In recent years photonic lattices have been used to study fundamental solid state phenomena including Bloch oscillations \cite{Bloch1928, Pertsch1999}, dynamic localisation \cite {Dunlap1986, Longhi2006, Szameit2010}, Bloch-Zener dynamics \cite {Dreisow2009}, and Landau-Zener dynamics \cite {2Dreisow2009}. These phenomena are each related to the manner in which a charged particle behaves in a periodic potential and external electric field. In such a system, a static electric field destroys the translational symmetry of the lattice, and the delocalised Bloch states become localised in space. These states are known as Wannier-Stark (W-S) states, originally predicted by G. H. Wannier in 1960 \cite{Wannier1960}. 

In the absence of an external electric field, the eigenstates of an electron in a periodic potential are the Bloch states. 
A static electric field ($\mathcal{E}_{dc}$) destroys the degeneracy of these spatially delocalised Bloch states. In this situation, the eigenstates and eigenfunctions are \cite{Fukuyama1973,Emin1987}
\begin{eqnarray}
\phi_m &=&\sum_n J_{n-m}(2\kappa/e\mathcal{E}_{dc}a) \vert n\rangle, \label{1} \\
E_m &=&(e\mathcal{E}_{dc}a)m \label{2},
\end{eqnarray}
where $J_{\nu}(\cdot)$ is the Bessel function of order $\nu$, $a$ is the lattice spacing, $4\kappa$ is the bandwidth, $e$ is the electronic charge and $\vert n\rangle$ are the Wannier states.  The span of the first Brillouin zone is $-\pi/a\le k \le \pi/a$. In the limit $2\kappa/e\mathcal{E}_{dc}a \rightarrow 0$ there is only one term in Eq. (\ref{1}),  i.e. the eigenstates exactly correspond to the localized Wannier states $\vert m\rangle$.
In fact, in a strong external electric field and weak inter-site interaction i.e. $2\kappa/e\mathcal{E}_{dc}a \ll 1$, the spatial width of W-S state is less than the inter-site separation, $a$. In this limit, the W-S states will be localised to a single lattice site, indicating that the energy offset $\Delta\!=\!e\mathcal{E}_{dc}a\gg \kappa$ generated by the static electric field inhibits the bare hopping between neighboring sites. Importantly, when driving the system, the strongly localized electronic states on individual lattice sites can interact with photons and tunnel to the nearest lattice sites. This type of tunneling with discrete energy exchange is known as photon-assisted tunneling; it has been observed in superconducting diodes\cite{Tien1963}, semiconductor superlattices \cite{Guimaraes1993}, quantum dots \cite{Kouwenhoven1994}, and also with cold atoms trapped in optical lattices \cite{Sias2008}.

\begin{figure}
\includegraphics[width=8.5cm]{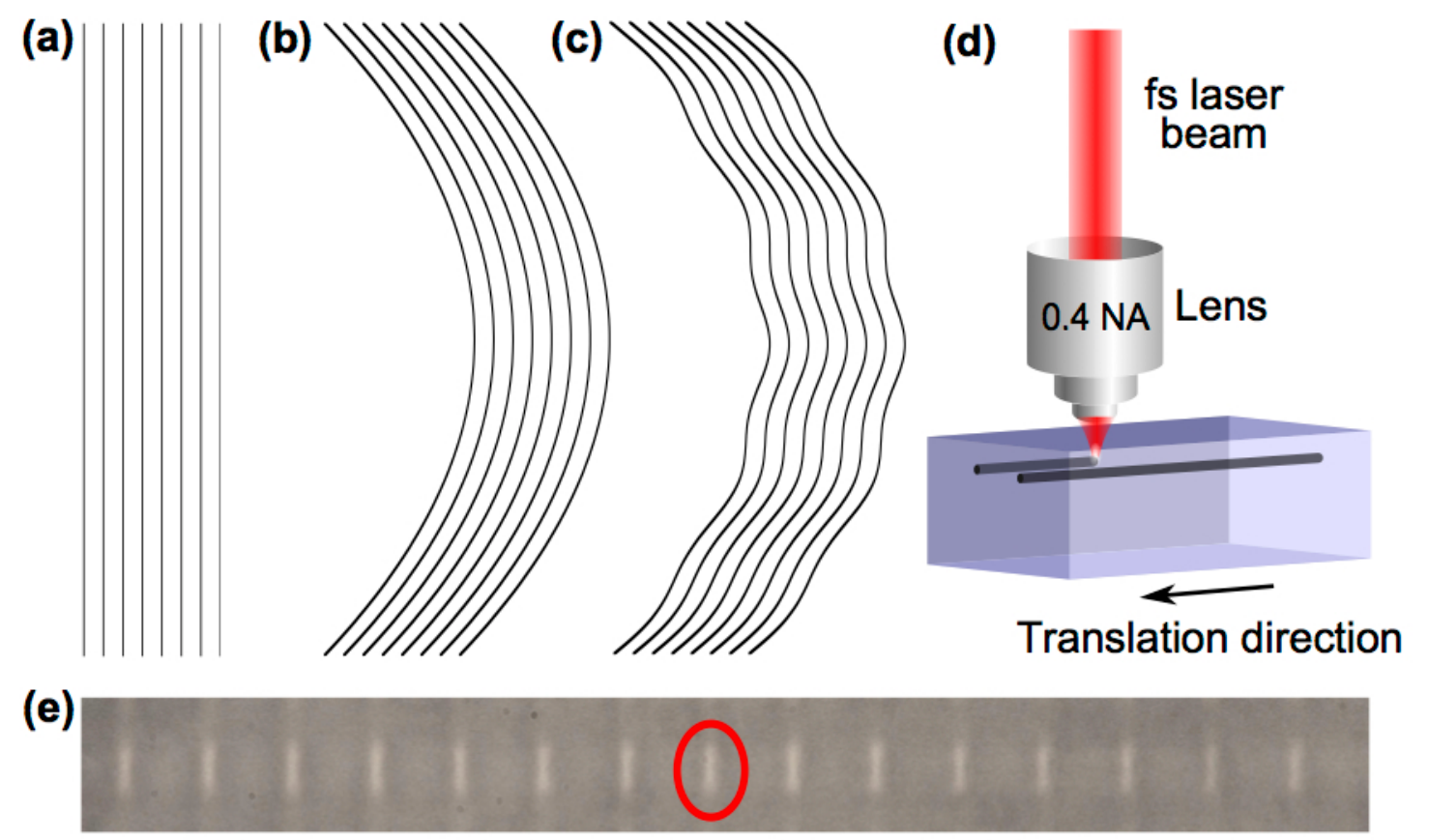}
\caption{Schematic diagrams of (a) an array of straight optical waveguides, (b) an array of circularly curved optical waveguides, (c) an array of circularly curved optical waveguides with a sinusoidal modulation. (d) Schematic diagram of the ultrafast laser inscription technique. (e) White-light transmission micrograph of the facet of an array of straight waveguides. Lattice constant $a$ = 16 $\mu m$. Each waveguide is single-mode at 780 nm wavelength. The central waveguide, indicated with the red circle, was excited for all measurements.
}
\label{fig1}
\end{figure}

In this Letter, we use photonic lattices, fabricated using the technique of ultrafast laser inscription , 
as a powerful platform to investigate the dynamics of W-S states in a strong static electric field and weak inter-site interaction. We demonstrate strong localization of the W-S state, seen for the first time using curved photonic lattices, where the curvature is analogous to the inverse of a static electric field in the electronic case.  When the electric field exceeds a threshold value, we observe that the W-S state becomes localised to a single lattice site. Importantly, we then also demonstrate that a strongly localised W-S state becomes delocalised when an appropriate (specific frequencies and amplitudes) sinusoidal modulation is applied to the lattice. The latter result constitutes the first photonic-crystal analogue of photon-assisted tunneling, based on fabricated sinusoidal modulations, offering a promising method for the generation of gauge fields in photonic lattices.

\textit{The photonic lattice.} The propagation of the electric field envelope $\tilde\Phi$ in the material is governed by 
\begin{equation}
i\lambdabar \frac {\partial \tilde\Phi}{\partial z}  =\Big[-\frac{\lambdabar^2}{2n_0} \nabla_\bot^2+V(x'-x_0(z),y)\Big]\tilde\Phi \label{neweq}
\end{equation}
where $\tilde\Phi$ depends on $x',y$ and $z$, $\lambda=2\pi\lambdabar$ is the free-space wavelength and $\nabla_\bot^2=\frac {\partial^2 }{\partial x'^2}+\frac {\partial^2 }{\partial y^2}$. $V(x',y)=\sum_nV_0(x'-x_n,y-y_m)$ describes the refractive index modulation in the transverse cross section where $V_0(x',y)$ is the refractive index profile of a single waveguide at position $x_n,y_n$. The function $x_0(z)$ determines the transverse shift of the whole lattice depending on the propagation distance $z$. By making a change of reference frame $x=x'-x_0(z)$ equation (\ref{neweq}) can be rewritten as
\begin{equation}
i\lambdabar \frac {\partial \Phi}{\partial z} =\Big[-\frac{\lambdabar^2}{2n_0} \nabla_\bot^2+V(x,y)-Fx \Big]\Phi
\end{equation}
with $F=-n_0\partial_z^2 x_0(z)$ and where $\Phi(x,y,z)$ is the consequently transformed state \cite {Heiblum1975, Longhi2006}. The transport of light in a circularly curved one-dimensional photonic lattice, with sinusoidal modulation (Fig \ref{fig1}(c)) is then governed by the paraxial equation \cite {Heiblum1975, Longhi2006}
\begin{eqnarray}
i\lambdabar \frac {\partial \Phi}{\partial z} & =&\Big[-\frac{\lambdabar^2}{2n_0} \nabla_\bot^2-\Delta n(x,y)-\frac{n_0}{R}x  \nonumber \\
&&  -n_0 A \omega_0^2 \sin(\omega_0 z)  x \Big] \Phi, \label{3}
\end{eqnarray}
where the lattice is bending along the $x$ direction with a radius of curvature $R$. The amplitude and frequency of the $z$-dependent "ac" modulation are $A$ and $\omega_0$ respectively, $n_0$ is the refractive index of the substrate material and  $\Delta n(x,y)$ is the transverse refractive index profile. Eq. (\ref{3}) is analogous to the Schr\"odinger equation of a particle with effective mass $n_0$ and charge $e$ moving in a 1D periodic potential $V(x)=-\Delta n(x)$, with an external (artificial) electric field $\mathcal{E}=\mathcal{E}_{dc}+\mathcal{E}_{ac}$, where $e\mathcal{E}=n_0/R+n_0A\omega_0^2  \sin(\omega_0 z)$. Here, $z$ plays the role of time,  $e\mathcal{E}_{dc}=n_0/R$, and $e\mathcal{E}_{ac}=n_0 A \omega_0^2 \sin(\omega_0 z)$ (see Eqs. (\ref{1}) and (\ref{2})). 

\begin{figure} 
\includegraphics[width=8.5cm]{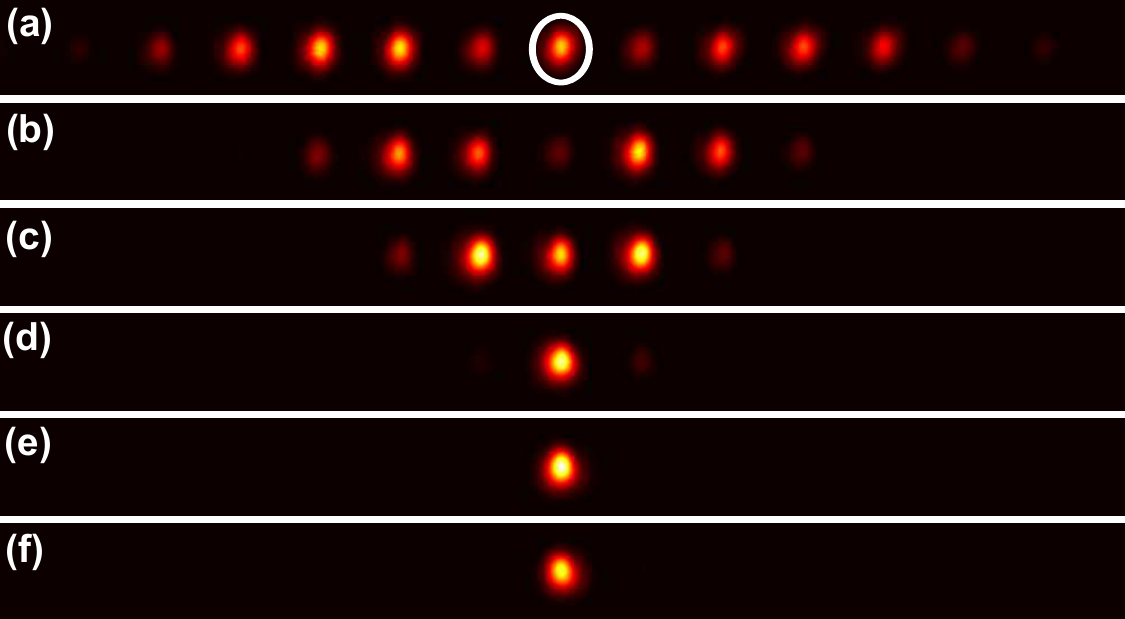}
\caption{Intensity distribution at the output of 30-mm-long circularly curved 1D lattices with radii of curvature (a) $R=\infty$ (i.e. straight lattice), (b) $R=1.5$ m, (c) $R=1.2$ m, (d) $R=0.5$ m, (e) $R=0.3$ m and (f) $R=0.2$ m. In each case, light was launched into the central waveguide.
}\label{fig2}
\end{figure}

For well-confined single-mode waveguides, Eq. (\ref 3) can be solved using the tight-binding approximation. When only the lowest band is excited, Eq. (\ref 3) gives the coupled-mode equations
\begin{equation}
i\frac{\partial\Phi_s}{\partial z}=-\kappa(\Phi_{s+1}+\Phi_{s-1})-s(\alpha+K\sin(\omega_0 z)) \Phi_{s} \label{4},
\end{equation}
where $\Phi_{s}$ is the electric field amplitude in the $s$:th waveguide, $\kappa$ is the nearest-neighbour coupling constant along the x-axis, $\alpha=n_0a/R\lambdabar$ is the strength of the external linear potential and $K=n_0 A \omega_0^2 a / \lambdabar$ is the strength of the ac modulation.
When only the dc field is present, and $2\kappa R \lambdabar/n_0a \ll 1$, the state excited at the input becomes localised to one site. 
When adding the ac field, this localised state can interact  with its neighbouring sites when the resonance condition
\begin{equation}
\omega_0 \nu = \alpha = n_0a/R\lambdabar, \quad \nu \in \mathbb{Z} , \label{5} 
\end{equation}
 is satisfied. This leads to an effective tunneling given by \cite{Eckardt2005}
\begin{equation}
\frac{\kappa_{\text{eff}}}{\kappa} =\bigg|J_{\nu}\bigg(\frac{K}{\omega_0}\bigg)\bigg|, \label{6}
\end{equation}
where $J_{\nu}$ is the Bessel function of order $\nu$. 

\begin{figure}
\centering
\includegraphics[width=1\linewidth]{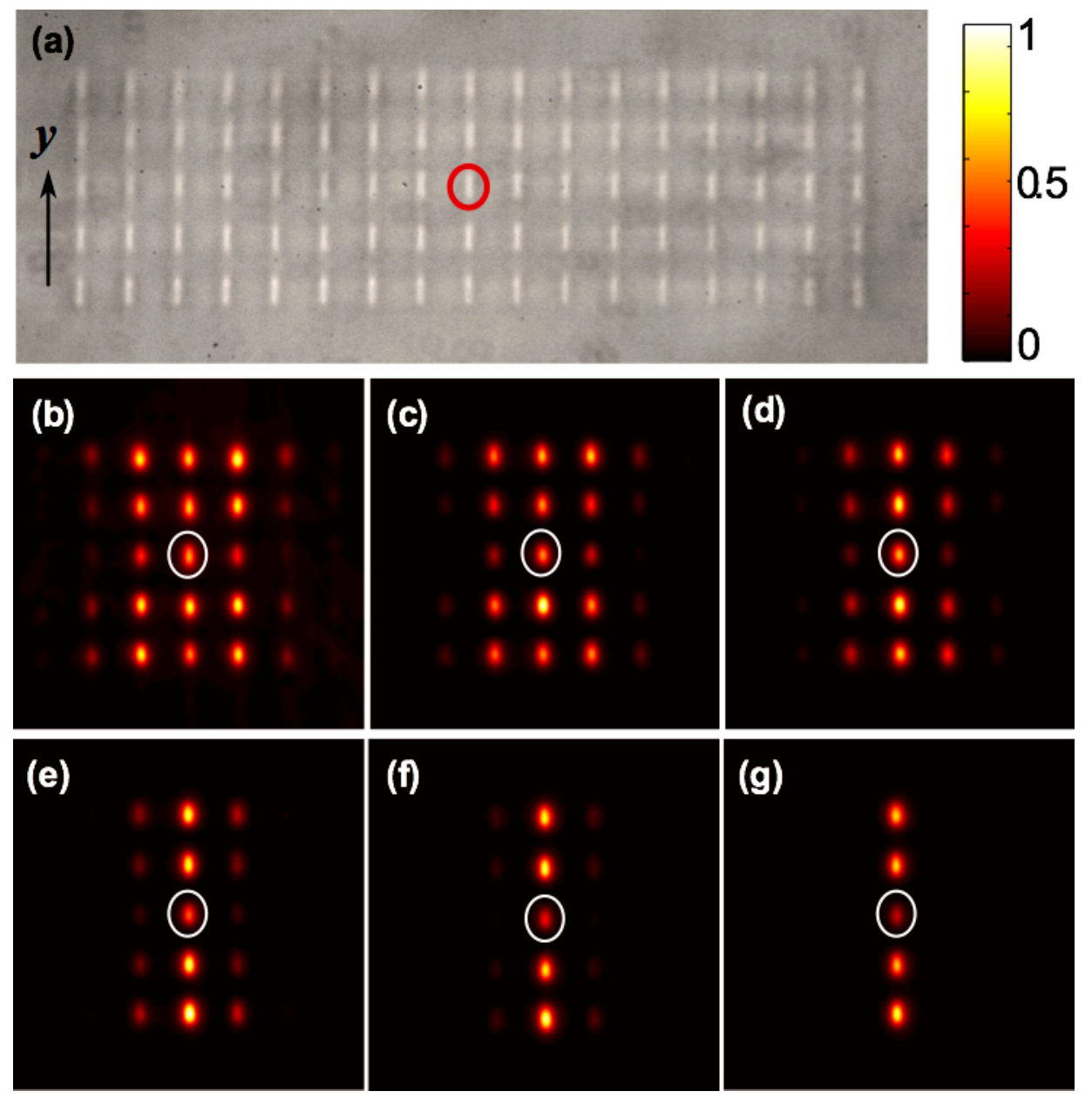}
\caption{(a) Micrograph of the facet of a square lattice. For all measurements, light was launched into the circled waveguide. (b)-(g) Intensity distribution at the output of 10-mm-long circularly curved square lattice with radii of curvature  (b) $R=\infty$ (i.e. straight lattice), (c) $R=1$ m, (d) $R=0.7$ m, (e) $R=0.5$ m, (f) $R=0.4$ m and (g) $R=0.3$ m. $\kappa_x=0.085$, $\text{mm}^{-1}$,  $\kappa_y=0.095$ $\text{mm}^{-1}$,  $\kappa_N=0.019$ $\text{mm}^{-1}$, $a$=15 $\mu$m.}
\label{fig3}
\end{figure}

\textit{Wannier-Stark localisation} To investigate Wannier-Stark localisation, fifteen 1D lattices (lattice constant $a$ = 16 $\mu$m) were fabricated using 
ultrafast laser inscription, Fig. \ref{fig1}(d). In these lattices, no modulation was created ($A\!=\!0$), however the radius of curvature of the lattice was varied between 1.5 and 0.1 m ($R=1.5, 1.4,..., 0.1$ m), see Fig. \ref{fig1}(b). An additional straight lattice was also fabricated ($R=\infty$), see Fig. \ref{fig1}(a). The white-light transmission micrograph of the facet of a lattice is shown in Fig. \ref{fig1}(e). The refractive index profile of each waveguide was controlled using the ``slit-beam" shaping method \cite{Ams2005}. Each waveguide was inscribed by translating the 30-mm-long glass sample (Corning Eagle$^{2000}$), at a translation speed of 8 mm-s$^{-1}$, once through the focus of a 500 kHz train of 1030 nm femtosecond laser pulses. The  laser inscription parameters were optimized to produce waveguides that were single-mode and well confined at 780 nm. For a more detailed description of the waveguide fabrication procedure, see Ref. \cite{Mukherjee2014}. 
\begin{figure}
\centering
\includegraphics[width=8.5cm]{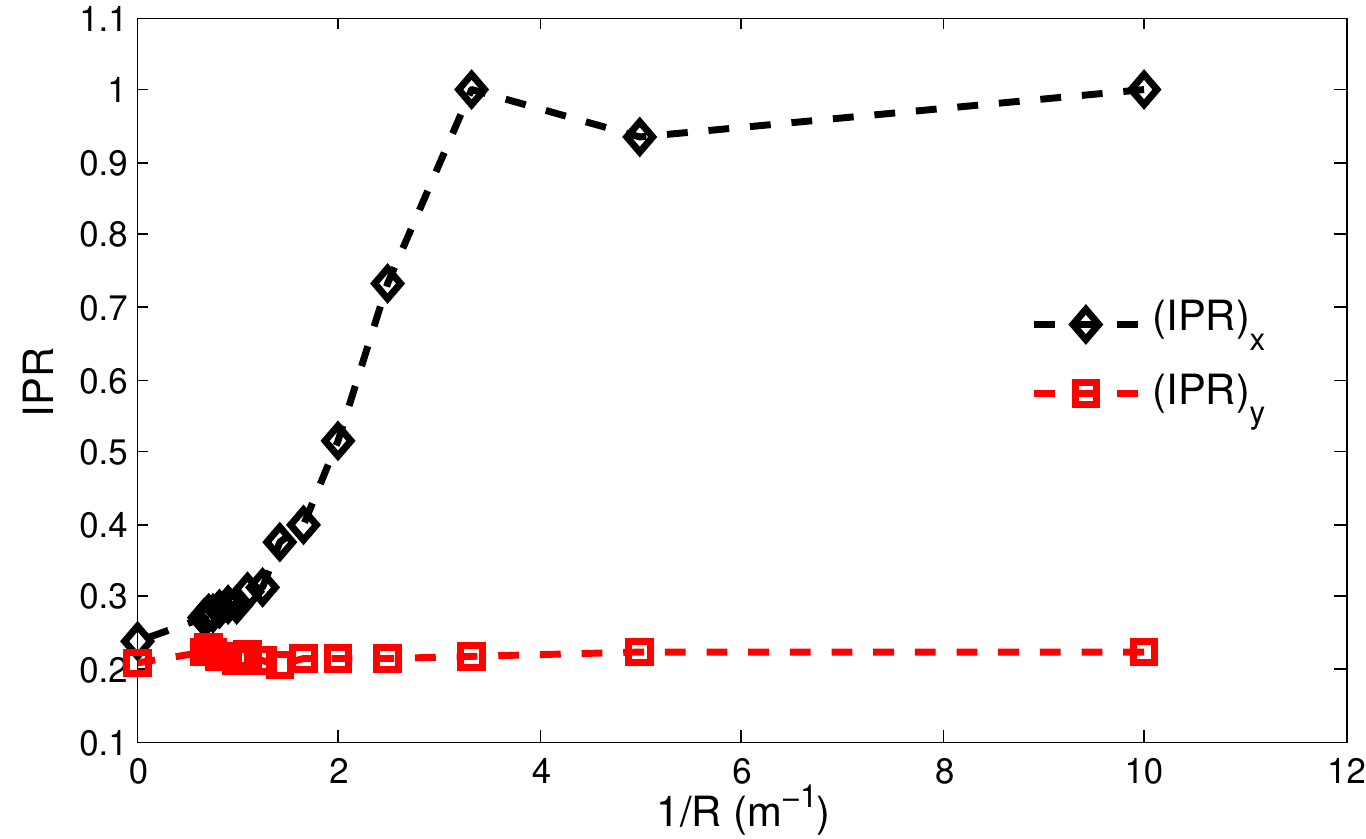}
\caption{Inverse participation ratio (IPR) is a measure of localisation and is defined as the average of the absolute value of the fourth power of the wave function. 
The IPRs along the x and y directions, (IPR)$_x$ and (IPR)$_y$ respectively, have been plotted as a function of the inverse of curvature radius $R$ which is a measure of the strength of dc electric field. There is no effect of electric field along the $y$ direction. Complete localisation (IPR=1) is observed in the $x$ direction as the electric field exceeds a threshold value.}
\label{fig4}
\end{figure}

The nearest-neighbour coupling $\kappa$ was measured to be 0.072 $\text{mm}^{-1}$. Fig. \ref{fig2} shows the output intensity distribution measured for the lattices with radii of curvature $R=\infty$ (i.e. the straight lattice), 1.5 m, 1.2 m, 0.5 m, 0.3 m and 0.2 m. It is clear from Fig. \ref{fig2} that the light becomes increasingly localized as the radius of curvature is reduced, as would be expected from Eq. (\ref{1}). To investigate this phenomenon further, we fabricated 10-mm-long 2D lattices with a lattice constant $a$ = 15 $\mu m$ along both $x$ and $y$ axes, where each lattice curves only along the $x$ direction. For these lattices, the measured coupling strengths along the $x$ and $y$ axes were 0.085 mm$^{-1}$ $(\kappa_x)$ and 0.095 mm$^{-1}$ $(\kappa_y)$ respectively. From simulations, the estimated value of next-nearest-neighbour coupling was 0.019 mm$^{-1}$ $(\kappa_N)$. As can be seen from Fig \ref{fig3}, localisation occurs only along the $x$ axis, the direction of the artificial electric field. To quantify localisation along the two axes, the inverse participation ratio (IPR) was calculated. The IPR is a measure of localisation and is defined as the inverse of the absolute value of the average of the fourth power of the wave function. For our purpose, the IPR for the $x$ axis was obtained by summing all the intensity values in each column to obtain a vector of values along the $x$ axis. The IPR was then calculated using this vector. The IPR along the $y$ axis was calculated using the same procedure, but by summing the rows rather than columns. For a localised state, the IPR is equal to 1. As can be seen from Fig. \ref{fig4}, there is no effect of electric field along the $y$ axis, as would be expected, but complete localisation 
is observed along the $x$ axis once the artificial electric field exceeds a threshold value. 

The localization phenomenon can also be explained using the theory of waveguide optics. It can be shown \cite {Heiblum1975}, using a conformal transformation, that a 1D array of circularly curved waveguides with periodic transverse refractive index profile is equivalent to an array of straight waveguides with a new refractive index profile. In the limit $a/R\ll 1$, the new refractive index profile is the superposition of the original periodic index profile and a linear ramp of refractive index, and the radius of curvature controls this ramp. In other words, the mode supported by each waveguide in the curved array has a different propagation constant $\beta$. As $R$ decreases, the difference in $\beta$ increases, resulting in partial transfer of light into the nearest waveguides via evanescent coupling. After a threshold value of $R$, there is no significant coupling between the nearest waveguides resulting in a complete spatial localization.

\begin{figure}
\centering
\includegraphics[width=8.5cm]{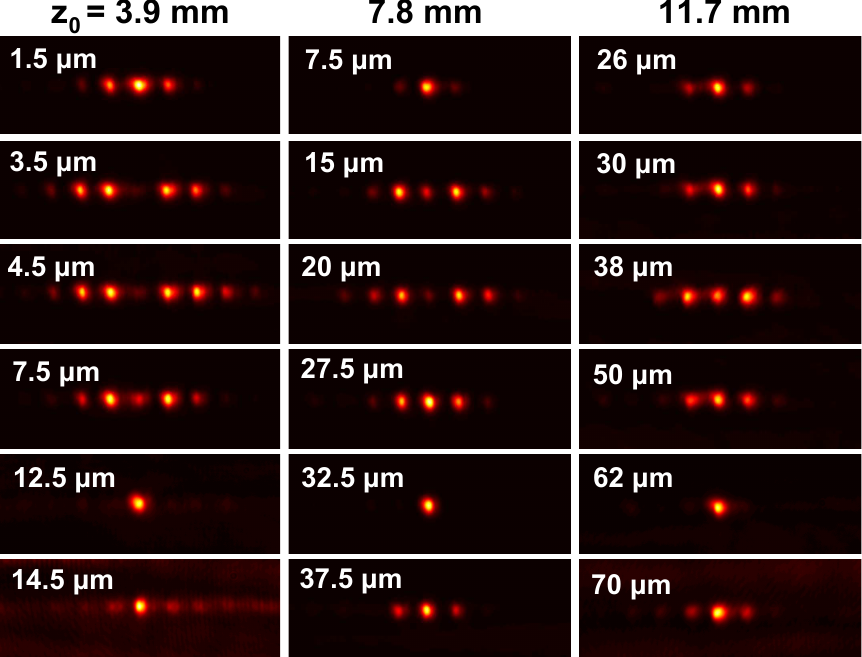}
\caption{Intensity distribution at the output of circularly curved lattices with sinusoidal modulation. Spatial periods of oscillation are 3.9 mm (1:st column), 7.8 mm (2:nd column) and 11.7 mm (3:rd column). For each period the amplitude of oscillation, $A$, was varied as shown. See also Fig. (\ref{fig6}).
}
\label{fig5}
\end{figure}

\begin{figure}
\centering
\includegraphics[width=8.5cm]{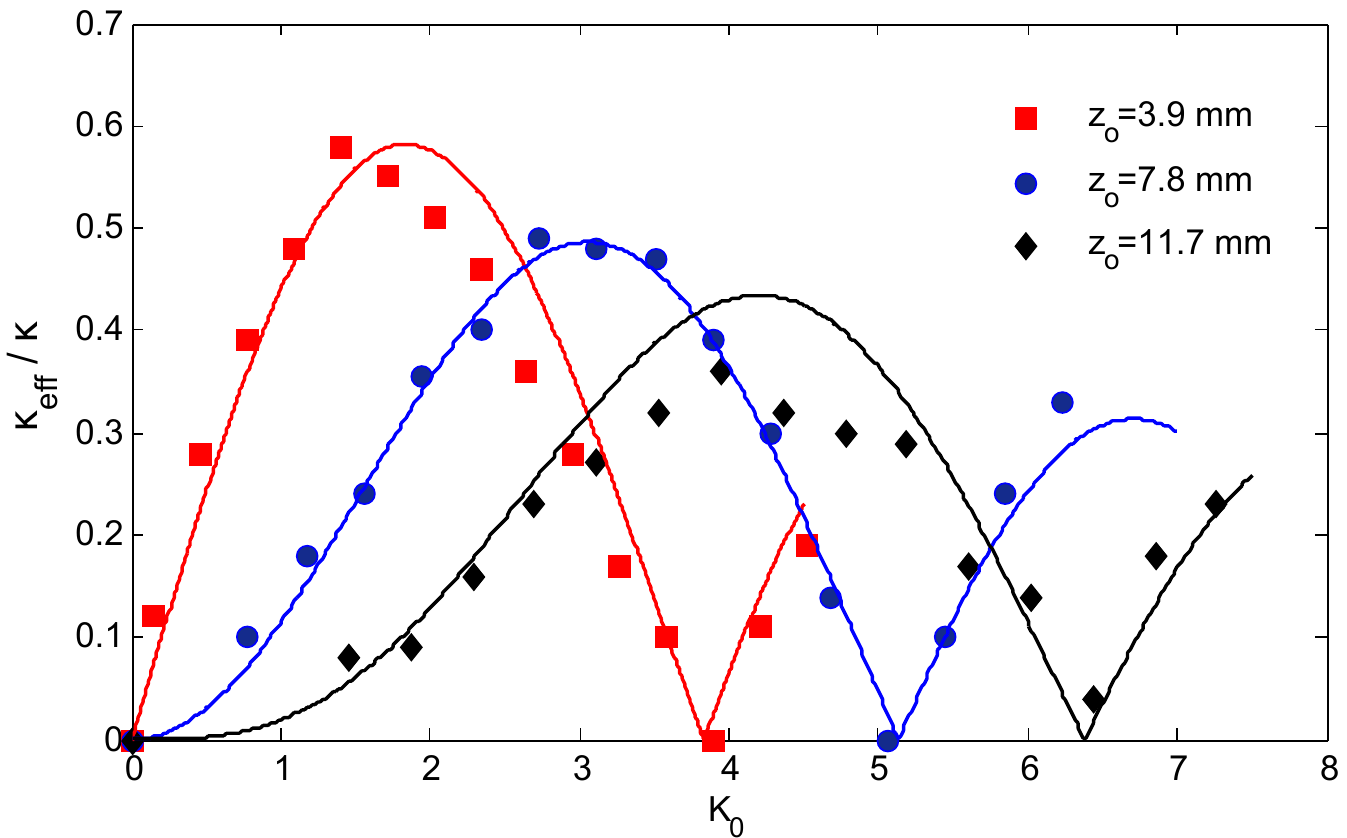}
\caption{ Graphical representation of  Fig. (\ref{fig5}). Variation of effective coupling with $K_0=K/\omega_0$. The solid lines are the absolute values of Bessel functions $|J_1|, |J_2|$ and $|J_3|$ respectively, which were predicted theoretically from Eq. (\ref{6}).}
\label{fig6}
\end{figure}

{\it Analog photon-assisted tunneling} To observe the effect analog to photon-assisted tunneling, three sets of 30-mm-long lattices were fabricated with a sinusoidal modulation (Fig. \ref{fig1}(c)). For sets 1, 2 and 3, the periods $z_0=2\pi/\omega_0$ were set to 3.9 mm, 7.8 mm and 11.7 mm respectively, corresponding to $\nu = 1, 2$ and 3 in Eq. (\ref{5}). For all sets, the radius of curvature and inter-site separation were set to $R$=120 mm and $a$=16 $\mu$m respectively. For each set, 15 lattices were fabricated, and the amplitude of oscillation, $A$, was varied; see Fig.  \ref{fig5}. The measured output intensity distributions are shown in Fig \ref{fig5}. The ratio of the effective coupling $\kappa_{\text{eff}}$ and the coupling strength for the array  of straight waveguides $\kappa$ was determined by comparing the output intensity distribution for the modulated curved lattice (Fig. \ref{fig1}(c)) and that for the straight lattice (Fig. \ref{fig1}(a)). The effective tunneling is plotted graphically as a function of $K_0=K/\omega_0$ in Fig \ref{fig6}, where it can be seen that the effective coupling rate has a characteristic (Bessel-function) dependency on $K_0$, as predicted by Eq. \ref{6}. This is clear evidence that the  tunneling has been partially restored through an analogue of photon-assisted tunneling. As a final note, it should be stressed that significant tunneling was absent when $z_0$ was not an integer multiple of 3.9 mm.

{\it Conclusion} In this Letter, we have demonstrated that an appropriately designed array of evanescently coupled curved optical waveguides can be used to observe a W-S state that is fully localized on a single lattice site. From the perspective of solid state physics, the localization is due to an analogue of a strong external dc electric field that breaks the degeneracy of the Bloch states and results in a Wannier-Stark ladder. We also demonstrate that tunneling in such photonic lattices can be restored by applying an analogue of an ac electric field, and that the strength of this tunneling obeys a characteristic dependency on the frequency and amplitude of the ac modulation, which is in excellent agreement with the existing theory of photon-assisted tunneling. By further tuning the spatial dependence of the laser-fabricated modulation, this method could be used to produce effective magnetic fluxes  \cite{Bermudez:2011prl,Kolovsky:2011,Goldman:2015pra} in 2D photonic lattices. The interplay between such artificial fields and the presence of non-linearities opens a promising route for the study of interacting particles in large magnetic fields.

{\it Acknowledgments} R.R.T. gratefully acknowledges funding from the UK Science and Technology Facilities Council (STFC) in the form of an STFC Advanced Fellowship (ST/H005595/1) and through the STFC Project Research and Development (STFC-PRD) scheme (ST/K00235X/1). RRT also thanks the European Union for funding via the OPTICON Research Infrastructure for Optical/IR astronomy (EU-FP7 226604). A.S. acknowledges support from the EPSRC CM-DTC. S.M. thanks Heriot-Watt University for a James Watt Ph.D Scholarship. N.G. is financed by the FRS-FNRS Belgium and by the  Belgian Science Policy Office under the Interuniversity Attraction Pole project P7/18 DYGEST. We acknowledge helpful discussions with Manuel Valiente.


\end{document}